\documentclass[12pt]{article}
\usepackage{amsfonts,amssymb}
\def\dac{\displaystyle\frac}

\def\d{\partial}
\def\[{\left[}
\def\]{\right]}
\def\({\left(}
\def\){\right)}
\def\p{\phantom}

\newcommand{\arctanh}{\mathop{\rm arctanh}\nolimits}
\newcommand{\diag}{\mathop{\rm diag}\nolimits}
\newcommand{\const}{\mathop{\rm const}\nolimits}
\begin{document}

\begin{center}
\Large\textbf{Accelerating Cosmologies in Lovelock Gravity with
Dilaton}\normalsize

I. V. Kirnos$^\dag$\footnote{e-mail: ikirnos@mail.ru}, A. N.
Makarenko$^\ddag$\footnote{e-mail: andre@tspu.edu.ru}

$^\dag$\textit{Tomsk State University, 634050, Russia, Tomsk, Lenin
prosp., 36}

$^\ddag$\textit{Tomsk State Pedagogical University, 634041, Russia,
Tomsk, Komsomol'sky prosp., 75}

\end{center}

\begin{abstract}
For the description of the Universe expansion, compatible with
observational data, a model of modified gravity --- Lovelock gravity
with dilaton --- is investigated. D-dimensional space with 3- and
(D-4)-dimensional maximally symmetric subspaces is considered. Space
without matter and space with perfect fluid are under test. In
various forms of the theory under way (third order without dilaton
and second order --- Einstein-Gauss-Bonnet gravity --- with dilaton
and without it) stationary, power-law, exponential and
exponent-of-exponent form cosmological solutions are obtained. Last
two forms include solutions which are clear to describe accelerating
expansion of 3-dimensional subspace. Also there is a set of
solutions describing cosmological expansion which does not tend to
isotropization in the presence of matter.
\end{abstract}

\section*{Introduction}

At present time there are numerous observational data known to be
incompatible with Standard Cosmological Model. On the one hand,
accelerating expansion observations from supernovae type Ia
\cite{kowalski} and gravitational lensing \cite{bernstein} allow us
to calculate metric tensor. On the other hand, evaluating of amount
of visible matter, energy-momentum tensor can be obtained. However,
it is impossible to satisfy Einstein equations by plugging in these
values. Then there are two possibilities: there is a great amount of
invisible matter or Einstein equations is not true. These
possibilities point out two approaches to the problem: to develop
theories of dark matter and dark energy or to modify theory of
gravity.

In the present article we are engaged in the second approach. At
that we will give attention exclusively to obtaining cosmological
acceleration in the space with extra dimensions or without them,
with perfect fluid or without matter, and to obtaining cosmological
solutions which do not tend to isotropization. The latter are
important for the reason that extra dimensions should be small and
then they should not expand such as visible ones. We will not deal
with gravitational lensing, galaxy moving in clusters, rotating
curves of galaxies and so on in the framework of theory under
investigation. Also we are not dealing with any issues related to
quantization.

Modified gravity has its beginning in 1920-th. The most popular
theories are Brans-Dicke theory \cite{brans-dicke, dicke}, Lovelock
gravity \cite{lovelock} and $f(R)$-gravity (see, e. g.
\cite{nojiri-odintsov}). However, for a long time these theories were not useful
for explanation of experimental data incompatible with general
relativity.

Here we will investigate a scalar-tensor extension of Lovelock
gravity --- Lovelock gravity with dilaton which might have its
origin in low-energy limit of string theory. Lovelock gravity with
dilaton contains scalar field $\varphi$ (dilaton), metric tensor
$g_{\mu\nu}$, matter fields $\Phi^I$ and is described (in
$D$-dimensional space-time) by Lagrangian
\begin{equation}\label{lagrangian} {\mathcal
L}=\sum_{p=1}^{m}2\alpha_p(\varphi)\delta^{\lambda_1\cdots\lambda_{2p}}_{\sigma_1\cdots\sigma_{2p}}
R_{\lambda_1\lambda_2}^{\phantom{\lambda_1\lambda_2}\sigma_1\sigma_2}R_{\lambda_3\lambda_4}^{\phantom
{\lambda_3\lambda_4}\sigma_3\sigma_4}\cdots
R_{\lambda_{2p-1}\lambda_{2p}}^{\phantom{\lambda_{2p-1}\lambda_{2p}}\sigma_{2p-1}\sigma_{2p}}
+g^{\mu\nu}\d_\mu\varphi\d_\nu\varphi-V(\varphi)+\mathcal{L}_M(\Phi^I,g_{\mu\nu}),\end{equation}
where $$ m=\frac1 2 D,\qquad\mbox{if $D$ is even,}$$ $$ m=\frac1 2
(D-1),\qquad\mbox{if $D$ is odd,}$$ $\alpha_p(\varphi)$,
$V(\varphi)$ are arbitrary functions of dilaton,
$\delta^{\mu_1\cdots\mu_k}_{\nu_1\cdots\nu_k}$ is the generalized
Kronecker delta and is equal to $1$ if $\nu_1\cdots\nu_k$ is even
transposition of $\mu_1\cdots\mu_k$, to $-1$ if odd one, and to zero
otherwise; $\mathcal{L}_M(\Phi^I,g_{\mu\nu})$ is the matter
Lagrangian. We shall call terms
\begin{equation}\label{lovelocklagrangian}{\mathcal L}_p=2\delta^{\lambda_1\cdots\lambda_{2p}}_{\sigma_1\cdots\sigma_{2p}}
R_{\lambda_1\lambda_2}^{\phantom{\lambda_1\lambda_2}\sigma_1\sigma_2}R_{\lambda_3\lambda_4}^{\phantom
{\lambda_3\lambda_4}\sigma_3\sigma_4}\cdots
R_{\lambda_{2p-1}\lambda_{2p}}^{\phantom{\lambda_{2p-1}\lambda_{2p}}\sigma_{2p-1}\sigma_{2p}}\end{equation}
as Lovelock Lagrangians of $p$-th order.

Theory under investigation is outstanding by the fact that its field
equations are nonlinear with respect to second derivatives of metric
tensor but does not involve higher derivatives.

Now say a few words on the researches in the framework of the theory
under consideration, which have already been done. Usually (see, e.
g. \cite{odintsov1, odintsov2, tomsk}) only 2-nd order (i. e.
$\alpha_p=0$ $\forall p>2$) of Lovelock gravity without dilaton
(so-called Einstein-Gauss-Bonnet gravity) is investigated. Solutions
for more complicated variants of the theory are not large in number.
 Investigations of third order Lovelock gravity (without dilaton)
can be found in \cite{dehghani/hole, dehghani/brane1,
dehghani/brane2}, studies of the second order with dilaton can be
found in \cite{odintsov3, bamba-guo-ohta}. Moreover, C.~C.~Briggs
obtains explicit formulae for the 4-th and 5-th Lovelock tensors.
Also we should draw attention to researches in
$f(R,\mathcal{L}_2)$-gravity \cite{odintsov3, odintsov4} and to
works in the 3-rd and 4-th orders of theory which is similar to
Lovelock gravity and obtained from string theory low-energy limit as
well as the latter \cite{nojiri-odintsov-sami, elizalde}.

In the present paper a set of solutions in second order with dilaton
and without it, also in third order without dilaton are obtained. In
the most of solutions extra spatial dimensions are assumed to exist.
Unobservability of them is explained by Kaluza-Klein approach (see,
e. g. \cite[p. 186]{carroll} and references therein) which is
briefly the following: extra dimensions are compactified on so small
scale that it is impossible to observe them (by present day device).

In the first section seven-dimensional third order Lovelock gravity
without dilaton is considered. Second section is devoted to the second
order with dilaton in spaces with various number of dimensions.

\section{Third order Lovelock gravity without dilaton}

Because it seems impossible to consider general case of Lovelock
gravity in a space with great number of dimensions, let us now
discuss third order Lovelock gravity without dilaton and without
cosmological constant: fields are $g_{\mu\nu}$ and $\Phi^I$,
Lagrangian is
$$\mathcal{L}_{Lovelock3}=R+\alpha_2
\mathcal{L}_2+\alpha_3\mathcal{L}_3+\mathcal{L}_M(\Phi^I,g_{\mu\nu}).$$
Here $\alpha_2$, $\alpha_3$ are constants. General expressions for
2-nd and 3-rd Lovelock Lagrangians are:
\begin{equation}\label{L2}\mathcal{L}_2=R_{\mu\nu\alpha\beta}R^{\mu\nu\alpha\beta}-4R_{\mu\nu}R^{\mu\nu}+R^2\end{equation}
is Gauss-Bonnet Lagrangian,
\begin{equation}\label{L3}\begin{array}{l}
\mathcal{L}_3=2R^{\mu\nu\sigma\kappa}R_{\sigma\kappa
\rho\tau}R^{\rho\tau}_{\phantom{\rho\tau}\mu\nu}+8R^{\mu\nu}_
{\phantom{\mu\nu}\sigma\rho}R^{\sigma\kappa}_{\phantom{\sigma
\kappa}\nu\tau}R^{\rho\tau}_{\phantom{\rho\tau}\mu\kappa}+
24R^{\mu\nu\sigma\kappa}R_{\sigma\kappa\nu\rho}R^{\rho}_
{\phantom{\rho}\mu}+\vphantom{\dac{a}{\dac{a}{a}}}\\ \qquad
{}+3RR^{\mu\nu\sigma\kappa}R_{\mu\nu\sigma\kappa}
+24R^{\mu\nu\sigma\kappa}R_{\sigma\mu}
R_{\kappa\nu}+16R^{\mu\nu}R_{\nu\sigma}R^{\sigma}_
{\phantom{\sigma}\mu}-12RR^{\mu\nu}R_{\mu\nu}+R^3
\end{array}\end{equation} is third Lovelock Lagrangian.

By variation of action $S=\int\mathcal{L}_{Lovelock3}d^Dx$ one may
obtain
\begin{equation}\label{our/equation} G^{(1)}_{\mu\nu}
+\alpha_2G^{(2)}_{\mu\nu}+\alpha_3G^{(3)}_{\mu\nu}=\dac{8\pi
G}{c^4}T_{\mu\nu}, \end{equation} where
\begin{equation}\label{G1} G^{(1)}_{\mu\nu}=R_{\mu\nu}-\frac1 2 R g_{\mu\nu}, \end{equation}
\begin{equation}\label{G2} G^{(2)}_{\mu\nu}=2(-R_{\mu\sigma\kappa\tau}
R^{\kappa\tau\sigma}_{\phantom{\kappa\tau\sigma}\nu}-2R_{\mu\rho\nu\sigma}R^{\rho\sigma}-
2R_{\mu\sigma}R^\sigma_{\p{\sigma}\nu}+RR_{\mu\nu})-\frac1 2
(R_{\rho\sigma\alpha\beta}R^{\rho\sigma\alpha\beta}-4R_{\alpha\beta}R^{\alpha\beta}+R^2)g_{\mu\nu},\end{equation}
and $G^{(3)}_{\mu\nu}$ is written down according to \cite{briggs4}:
\begin{equation}\label{G3}\begin{array}{l}
G^{(3)}_{\mu\nu}=\dac{1}{2}\biggl(\left.-g_{\mu\nu}R^3+12g_{\mu\nu}RR_{\alpha\beta}R^{\alpha\beta}-
3g_{\mu\nu}RR_{\alpha\beta\sigma\kappa}R^{\alpha\beta\sigma\kappa}-
16g_{\mu\nu}R_{\alpha}^{\p{\alpha}\beta}R_{\beta}^{\p{\beta}\sigma}R_{\sigma}^{\p{\sigma}\alpha}+\vphantom{\dac{a}{\dac{a}{a}}}\right.\\
\quad{}\left.+
24g_{\mu\nu}R_{\alpha\beta}R_{\sigma\kappa}R^{\alpha\sigma\beta\kappa}+
24g_{\mu\nu}R_{\alpha}^{\p{\alpha}\beta}R^{\alpha\sigma\kappa\rho}R_{\beta\sigma\kappa\rho}+
2g_{\mu\nu}R_{\alpha\beta}^{\p{\alpha\beta}\sigma\kappa}R_{\sigma\kappa}^{\p{\sigma\kappa}\rho\lambda}
R_{\rho\lambda}^{\p{\rho\lambda}\alpha\beta}-\vphantom{\dac{a}{\dac{a}{a}}} \right.\\
\quad{}\left.-
8g_{\mu\nu}R_{\alpha\beta}^{\p{\alpha\beta}\sigma\kappa}R_{\sigma\rho}^{\p{\sigma\rho}\alpha\lambda}
R_{\kappa\lambda}^{\p{\kappa\lambda}\beta\rho}+
6R_{\mu\nu}R^2-24RR_{\mu}^{\p{\mu}\sigma}R_{\sigma\nu}-24R_{\mu\nu}R_{\alpha\beta}R^{\alpha\beta}+\vphantom{\dac{a}{\dac{a}{a}}}\right.\\
\quad{}\left.+
48R_{\mu}^{\p{\mu}\alpha}R_{\alpha}^{\p{\alpha}\beta}R_{\beta\nu}+
48R_{\mu}^{\p{\mu}\alpha}R^{\beta\sigma}R_{\alpha\beta\sigma\nu}+
6R_{\mu\nu}R_{\alpha\beta\sigma\kappa}R^{\alpha\beta\sigma\kappa}-
24R_{\mu\alpha}R_{\nu\beta\sigma\kappa}R^{\alpha\beta\sigma\kappa}+\vphantom{\dac{a}{\dac{a}{a}}}\right.\\
\quad{}\left.+
24RR_{\mu\sigma\nu\kappa}R^{\sigma\kappa}+12RR_{\mu\alpha\beta\sigma}R_{\nu}^{\p{\nu}\alpha\beta\sigma}-
48R_{\mu\alpha\nu\beta}R_{\sigma}^{\p{\sigma}\alpha}R^{\sigma\beta}-
48R_{\mu\alpha\beta\sigma}R_{\nu}^{\p{\nu}\beta}R^{\alpha\sigma}+\vphantom{\dac{a}{\dac{a}{a}}}\right.\\
\quad{}\left.+
48R_{\mu\alpha\nu\beta}R_{\sigma\kappa}R^{\alpha\sigma\beta\kappa}-
24R_{\mu\alpha\beta\sigma}R^{\kappa\alpha\beta\sigma}R_{\kappa\nu}-
24R_{\mu\alpha\beta\sigma}R^{\alpha\kappa}R_{\nu\kappa}^{\p{\nu\kappa}\beta\sigma}+\vphantom{\dac{a}{\dac{a}{a}}}\right.\\
\quad{}\left.+
48R_{\mu}^{\p{\mu}\alpha\beta\sigma}R_{\beta}^{\p{\beta}\kappa}R_{\sigma\kappa\nu\alpha}+
24R_{\mu\alpha\nu\beta}R^{\alpha}_{\p{\alpha}\sigma\kappa\rho}R^{\beta\sigma\kappa\rho}+
12R_{\mu}^{\p{\mu}\alpha\beta\sigma}R_{\beta\sigma}^{\p{\beta\sigma}\kappa\rho}R_{\kappa\rho\alpha\nu}+
\vphantom{\dac{a}{\dac{a}{a}}} \right.\\
\quad{}\left.+48R_{\mu\alpha}^{\p{\mu\alpha}\beta\sigma}R_{\beta\rho\nu\lambda}
R_{\sigma}^{\p{\sigma}\lambda\alpha\rho}\right.\biggr).
\end{array}\end{equation}

\subsection{Cosmological equations}

Now consider seven-dimensional\footnote{We elected 7-dimensional
space for two reasons. First, just in such a space nonzero third
order of Lovelock gravity arises. Second, in such a space we obtain
general exact solution of the present form for the second order
theory (see below).} flat space and assume metric tensor to get the
form
\begin{equation}\label{metric}g_{\mu\nu}=\diag\{-1,a^2(t),a^2(t),a^2(t),b^2(t),b^2(t),b^2(t)\}.\end{equation}
Furthermore, let $T_{\mu\nu}=0$.

From such a metric one can obtain nonzero Christoffel symbols:
$$\Gamma^0_{ii}=a\dot{a},\quad \Gamma^0_{aa}=b\dot{b}\quad
\Gamma^i_{i0}=\Gamma^i_{0i}=\dac{\dot{a}}{a},\quad
\Gamma^c_{c0}=\Gamma^c_{0c}=\dac{\dot{b}}{b}$$ (Latin indexes from
the middle of alphabet i,j,k,... run over visible subspace, and
Latin indexes from the beginning of alphabet a,b,c,... run over
extra subspace; index $0$ notice the time coordinate; Greek indexes
run over all the space). Nonzero components of Riemann tensor are
\begin{equation}\label{our/riemann}\begin{array}{ll}
R^{0}_{\p{0}i0i}=a(t)\ddot{a}(t);\qquad&
R^0_{\p{0}c0c}=b(t)\ddot{b}(t);\vphantom{\dac{a}{\dac{a}{a}}}\\
R^i_{\p{i}0i0}=-\dac{\ddot{a}(t)}{a(t)};\qquad& R^c_{\p{c}0c0}=-\dac{\ddot{b}(t)}{b(t)},\vphantom{\dac{a}{\dac{a}{a}}}\\
R^i_{\p{i}jij}=\dot{a}^2(t),\quad i\neq j;\qquad&
R^c_{\p{c}dcd}=\dot{b}^2(t),\quad c\neq d;
\vphantom{\dac{a}{\dac{a}{a}}}\\
R^i_{\p{i}cic}=\dac{\dot{a}(t)}{a(t)}b(t)\dot{b}(t);\qquad&
R^c_{\p{c}ici}=\dac{\dot{b}(t)}{b(t)}a(t)\dot{a}(t).\vphantom{\dac{a}{\dac{a}{a}}}\\
\end{array}\end{equation}

Now the field equations (\ref{our/equation}) are
\begin{equation}\label{system}\left\{\begin{array}{l}
H^2+3Hh+h^2+12\alpha_2 H^3 h+36\alpha_2 H^2 h^2 +12\alpha_2 H
h^3-240\alpha_3 H^3 h^3=0,\\ \\ \dot{H}\(2+24\alpha_2 H h+24\alpha_2
h^2-288\alpha_3 H h^3\)+\dot{h}\(3+48\alpha_2 H h+12\alpha_2
H^2+\right.\\ \left. \quad {}+12\alpha_2
h^2-432\alpha_3 H^2 h^2\)+3H^2+6Hh+6h^2+72\alpha_2 H^2 h^2+\\
\quad {}+72\alpha_2 H h^3+24\alpha_2 H^3 h +12\alpha_2 h^4
-432\alpha_3 H^2 h^4-288\alpha_3 H^3 h^3=0,\\ \\
\dot{H}(3+48\alpha_2 Hh+12\alpha_2 h^2+12\alpha_2 H^2-432\alpha_3
H^2h^2)+\dot{h}(2+24\alpha_2Hh+\\ \quad {}+24\alpha_2
H^2-288\alpha_3 H^3h)+3h^2+6H h+6H^2+72\alpha_2 H^2h^2+\\ \quad
{}+72\alpha_2H^3h+24\alpha_2Hh^3+ 12\alpha_2H^4-432\alpha_3H^4
h^2-288\alpha_3H^3h^3=0.\end{array}\right.
\end{equation}

Here
\begin{equation}\label{hubble}H(t)=\dot{a}(t)/a(t),\qquad
h(t)=\dot{b}(t)/b(t)\end{equation} are Hubble parameters for visible
and extra dimensions respectively.

Note that third equation is consequence of two other equations. A
reader is asked to verify this correlation by himself.

It seems reasonable to begin consideration of system (\ref{system})
with allowing $\alpha_2=\alpha_3=0$ and solving therefore
7-dimensional Einstein equations. But all solutions of them are only
particular cases of generalized Kasner solution (see \cite{kasner}
and \cite[\S 117]{landau-lifshitz} for 4-dimensional case).
Particularly, in our case (\ref{metric}) there is a solution
\begin{equation}\label{1solution2}\begin{array}{l}
a(t)=a^{(0)}\dac{1}{(t_0-t)^{\frac{3+\sqrt{5}}{6(\sqrt{5}+2)}}}\approx
a^{(0)}\dac{1}{(t_0-t)^{0.206}}, \\
b(t)=b^{(0)}(t_0-t)^{\frac{3-\sqrt{5}}{6(\sqrt{5}-2)}}\approx
b^{(0)}(t_0-t)^{0.539},\vphantom{\dac{l}{l}}\end{array}\end{equation}
where $a(t)$ describes accelerated expansion of visible subspace.

\subsection{General solution in the second order}

Assume $\alpha_3=0$ in (\ref{system}). Then we have
\begin{equation}\label{system2}\left\{\begin{array}{l}
H^2+3Hh+h^2+12\alpha_2 H^3 h+36\alpha_2 H^2 h^2 +12\alpha_2 H
h^3=0,\\ \\ \dot{H}\(2+24\alpha_2 H h+24\alpha_2
h^2\)+\dot{h}\(3+48\alpha_2 H h+12\alpha_2 H^2+\right.\\ \left.
\quad {}+12\alpha_2 h^2\)+3H^2+6Hh+6h^2+72\alpha_2 H^2 h^2+\\
\quad {}+72\alpha_2 H h^3+24\alpha_2 H^3 h +12\alpha_2 h^4=0,\\
\\ \dot{H}(3+48\alpha_2 Hh+12\alpha_2 h^2+12\alpha_2
H^2)+\dot{h}(2+24\alpha_2Hh+\\ \quad {}+24\alpha_2 H^2)+3h^2+6H
h+6H^2+72\alpha_2 H^2h^2+\\ \quad {}+72\alpha_2H^3h+
12\alpha_2H^4=0.\end{array}\right.
\end{equation}

From the first equation there are 3 possibilities:
$$H=-\dac{1}{12\alpha_2h},\qquad H=-\dac{3-\sqrt{5}}{2}h,\qquad
H=-\dac{3+\sqrt{5}}{2}h.$$ Second and third possibilities are
satisfied in all cases ($H>0$ if $h<0$), first one --- under
$\alpha_2>0$. Consider them one by one.

\textbf{1.} $H=-\dac{1}{12\alpha_2h}$. Plugging into the second
equation we have
$$\dot{h}=-\dac{1728\alpha_2^3h^6+1}{12\alpha_2(12\alpha_2h^2+144\alpha_2^2h^4+1)}.$$
Then $$h=\dac{1}{6}\sqrt{\dac{3}{\alpha_2}}x,$$ where $x$ obeys
equation
$$ x^5+3\gamma x^4-5x^3-5\gamma x^2 +3x+\gamma=0,$$ with
$$\gamma=\tan\(\dac{\sqrt{3}(t+t_0)}{2\sqrt{\alpha_2}}\).$$

\textbf{2.} $H=-\dac{3-\sqrt{5}}{2}h$. Plugging into the second
equation we have
$$\dot{h}=\dac{3h^2(-40\alpha_2h^2+16\alpha_2h^2\sqrt{5}-\sqrt{5}+5)}
{2(-\sqrt{5}+18\alpha_2h^2\sqrt{5}-30\alpha_2h^2)}.$$ Then $h$ obeys
equation
$$8\arctanh\(-\sqrt{\alpha_2}h+\sqrt{5\alpha_2}h\)\sqrt{\alpha_2}h
+1-6(t+t_0)h+\sqrt{5}=0.$$

\textbf{3.} $H=-\dac{3+\sqrt{5}}{2}h$. Plugging into the second
equation we have
$$\dot{h}=\dac{3h^2(40\alpha_2h^2-5-\sqrt{5}+16\alpha_2h^2\sqrt{5})}{2(30\alpha_2h^2+18\alpha_2h^2\sqrt{5}-\sqrt{5})}.$$
Then $h$ obeys equation
$$8\arctanh\(\sqrt{\alpha_2}h+\sqrt{5\alpha_2}h\)\sqrt{\alpha_2}h-1+6(t+t_0)h+\sqrt{5}=0.$$

In all cases above there is only parametric dependence $H(t)$ and
$h(t)$. Some explicit solutions in Einstein-Gauss-Bonnet gravity for
7 or other dimensions will be obtained in the second section.

\subsection{Exponential solution in the third order}

Now consider equations of the third order Lovelock gravity with
constant Hubble parameters: $$ \dot{H}=0, \qquad \dot{h}=0.$$

Then the system (\ref{system}) takes a form
\begin{equation}\label{system0}\left\{\begin{array}{l}
H^2+3Hh+h^2+12\alpha_2 H^3 h+36\alpha_2 H^2 h^2 +12\alpha_2 H
h^3-240\alpha_3 H^3 h^3=0,\\ \\ 3H^2+6Hh+6h^2+72\alpha_2 H^2
h^2+72\alpha_2 H h^3+24\alpha_2 H^3 h +12\alpha_2 h^4-\\ \qquad
\qquad \qquad \qquad \qquad \qquad \qquad \qquad \quad {}
-432\alpha_3 H^2 h^4-288\alpha_3 H^3 h^3=0,\\ \\ 3h^2+6H
h+6H^2+72\alpha_2 H^2h^2+72\alpha_2H^3h+24\alpha_2Hh^3+
12\alpha_2H^4-\\ \qquad \qquad \qquad \qquad \qquad \qquad \qquad
\qquad \quad {}-432\alpha_3H^4
h^2-288\alpha_3H^3h^3=0.\end{array}\right.
\end{equation}

Subtracting the second equation from the third one we have
$$\(H^2-h^2\)\[1+16\alpha_2 H h +4\alpha_2 \(H^2+h^2\)-144\alpha_3
H^2 h^2\]=0.$$ Then $h=-H$ is a solution of this equation.

Plugging this equality into the second equation of (\ref{system0})
one can obtain 4 various solutions: \begin{equation}\label{H1}
H=\pm\sqrt{\dac{-\alpha_2\pm
\sqrt{\alpha_2^2+12\alpha_3}}{24\alpha_3}}.\end{equation} Because of
the expansion of visible dimensions let us take only $H>0$. Then we
have
\begin{equation}\label{exponent}H=\sqrt{\dac{-\alpha_2\pm
\sqrt{\alpha_2^2+12\alpha_3}}{24\alpha_3}}, \qquad
h=-\sqrt{\dac{-\alpha_2\pm
\sqrt{\alpha_2^2+12\alpha_3}}{24\alpha_3}}.\end{equation}

Plugging $h=-H$ into the first equation of (\ref{system0}) we obtain
\begin{equation}\label{H2}H=\pm\sqrt{\dac{-\alpha_2\pm
\sqrt{\alpha_2^2+(20/3)\alpha_3}}{40\alpha_3}}.\end{equation}
Comparing that with (\ref{H1}) we have
$$\alpha_3=-\frac{1}{12}\alpha_2^2.$$ Then $$H=\dac{1}{\sqrt{2\alpha_2}},\qquad
h=-\dac{1}{\sqrt{2\alpha_2}}.$$

Therefore (because of (\ref{hubble})), \begin{equation}\label{main}
a(t)=C_1 \exp{\(\dac{t}{\sqrt{2\alpha_2}}\)},\qquad  b(t)=C_2
\exp{\(-\dac{t}{\sqrt{2\alpha_2}}\)},\end{equation} where $C_1$,
$C_2$ are arbitrary positive constants.

It is clear that $\dot a(t)>0,$ $\ddot a(t)>0$, i. e. the abovementioned solution
describes accelerated expansion of visible dimensions. At that extra
dimensions shrink. Then it is possible that visible and extra
dimensions were equivalent and the Universe look as 4-dimensional
one only after expansion of one subspace and contraction of another
one. Solution (\ref{main}) seems to be useful for the description of
inflation.

Finally, if $T_{\mu\nu}=0$, $\alpha_2>0$ and
$\alpha_3=-\dac{1}{12}\alpha_2^2$ then system (\ref{our/equation})
has solution (\ref{metric}) where functions $a(t)$ and $b(t)$ are
expressed by (\ref{main}) with arbitrary positive constants $C_1$
and $C_2$.

\section{Einstein-Gauss-Bonnet gravity with dilaton}

Such a theory is the following: fields $g_{\mu\nu}$, $\varphi$,
$\Phi^I$; Lagrangian
\begin{equation}\label{EGBd-lagr}\mathcal{L}_{EGBd}=R+g^{\mu\nu}\d_\mu\varphi\d_\nu\varphi-V(\varphi)+
\varepsilon(\varphi)\mathcal{L}_2+\mathcal{L}_M(\Phi^I,g_{\mu\nu}).\end{equation}

Here $\varepsilon(\varphi)$, $V(\varphi)$ are functions of dilaton
$\varphi$, $\mathcal{L}_2$ is a Gauss-Bonnet Lagrangian (\ref{L2}).
Theory under consideration is different from generalized
Brans-Dicke theory (see, e. g. \cite{carroll}) even in 4-dimensional
space, that is why we investigate both (3+1)-dimensional space and
spaces with extra dimensions (where Einstein-Gauss-Bonnet gravity
without dilaton is sensible).

Variating the action with Lagrangian (\ref{EGBd-lagr}) we get field
equations:
\begin{equation}\label{general-equations}\left\{
\begin{array}{l}
R_{\mu\nu}-\dac{1}{2}Rg_{\mu\nu}-\dac{1}{2}g_{\mu\nu}g^{\alpha\beta}
\d_\alpha\varphi\d_\beta\varphi+\dac{1}{2}g_{\mu\nu}V(\varphi)+\varepsilon(\varphi)G^{(2)}_{\mu\nu}+
\d_\mu\varphi\d_\nu\varphi-\\
\quad{}-\vphantom{\dac{\dac{1}{2}}{2}}
4\(R_{\mu\nu}-\dac{1}{2}Rg_{\mu\nu}\) \Box\varepsilon(\varphi)-
4\(R_{\mu\phantom{\alpha\beta}\nu}^{\phantom{\mu}\alpha\beta}+R^{\alpha\beta}
g_{\mu\nu}\)\nabla_\alpha\nabla_\beta\varepsilon(\varphi)+\\
\quad{}+
8R^\alpha_{\phantom{\alpha}\mu}\nabla_\alpha\nabla_\nu\varepsilon(\varphi)-
2R\nabla_\mu\nabla_\nu\varepsilon(\varphi)=\vphantom{\dac{\dac{1}{2}}{2}}\dac{8\pi
G}{c^4}T_{\mu\nu};\\
2\Box\varphi+V'(\varphi)-\vphantom{\dac{\dac{1}{2}}{2}}\varepsilon'(\varphi)\mathcal{L}_2=0.
\end{array}\right. \end{equation}
Here $G^{(2)}_{\mu\nu}$ is the second Lovelock tensor (\ref{G2}).

\subsection{Cosmological equations}

Consider space of $D=p+q+1$ dimensions with two maximally symmetric
subspaces: $p$-dimensional and $q$-dimensional. Square interval in
such a space is
\begin{equation}\label{solution-form}ds^2=-e^{2u_0(t)}dt^2+e^{2u_1(t)}ds_p^2+e^{2u_2(t)}ds_q^2,\end{equation}
where $ds_p^2$ and $ds_q^2$ are square intervals in $p$- and
$q$-dimensional subspaces respectively, $u_0(t), u_1(t), u_2(t)$ are
arbitrary functions of time $t$.

If metric is (\ref{solution-form}) then Christoffel symbols are (as
above, Latin indexes from the middle of alphabet i,j,k,... run over
visible $p$-subspace, and Latin indexes from the beginning of
alphabet a,b,c,... run over extra $q$-subspace; index $0$ notice the
time coordinate; Greek indexes run over all the space)
 \begin{equation}\label{Gamma}\begin{array}{c}\Gamma^0_{00}=\dot u_0,\qquad
\Gamma^0_{ij}=\dot u_1 e^{-2u_0}g_{ij},\qquad \Gamma^0_{ab}=\dot
u_2e^{-2u_0}g_{ab},\\
\Gamma^i_{jk}=\widetilde{\Gamma}^i_{jk},\qquad
\Gamma^i_{0i}=\Gamma^i_{i0}=\dot u_1,\qquad
\Gamma^a_{bc}=\widetilde{\Gamma}^a_{bc},\qquad
\Gamma^a_{0a}=\Gamma^a_{a0}=\dot u_2.\vphantom{\dac 1
2}\end{array}\end{equation}

Riemann tensor, Ricci tensor and scalar curvature are
$$\begin{array}{l}{R^0}_{i0j}=e^{-2u_0}Xg_{ij},\quad {R^0}_{a0b}=e^{-2u_0}Yg_{ab},\vphantom{\dac 1
2}\\ {R^i}_{jkl}=e^{-2u_0}A_p(\delta^i_k g_{jl}-\delta^i_l
g_{jk}),\quad {R^i}_{ajb}=e^{-2u_0}\dot{u}_1\dot{u}_2\delta^i_j
g_{ab},\vphantom{\dac 1 2}\\
{R^a}_{bcd}=e^{-2u_0}A_q(\delta^a_c g_{bd}-\delta^a_d g_{bc}),\quad
R_{00}=-pX-qY, \vphantom{\dac 1 2}\end{array}$$
$$R_{ij}=e^{-2u_0}\(X+(p-1)A_p+q\dot u_1\dot u_2\)g_{ij},\quad R_{ab}=e^{-2u_0}\(Y+(q-1)A_q+p\dot u_1\dot u_2\)g_{ab},$$
$$R=e^{-2u_0}\[2pX+2qY+p_1 A_p+q_1 A_q+2pq\dot u_1\dot u_2\].$$

Gauss-Bonnet Lagrangian (\ref{L2}):
$$\begin{array}{l}\mathcal{L}_2=e^{-4u_0}\vphantom{\dac{\dac{1}{2}}{2}}
\left\{p_3A_p^2+2p_1q_1A_pA_q+q_3A_q^2+\vphantom{\dac 1 2}
4\dot{u}_1\dot{u}_2(p_2qA_p+pq_2A_q)+\right.
4p_1q_1\dot{u}_1^2\dot{u}_2^2+\\
\quad{}+
4pX[(p-1)_2A_p+q_1A_q+2(p-1)q\dot{u}_1\dot{u}_2]+\vphantom{\dac{\dac{1}{2}}{2}}
\left.\vphantom{\dac 1 2}
4qY[p_1A_p+(q-1)_2A_q+2p(q-1)\dot{u}_1\dot{u}_2]\right\}
\vphantom{\dac{\dac{1}{2}}{2}}.\end{array}$$

Here we introduce the following notations:
\begin{equation}\label{signs1}\begin{array}{ll}A_p\equiv
\dot{u}_1^2+\sigma_p
e^{2(u_0-u_1)},& A_q\equiv\dot{u}_2^2+\sigma_q e^{2(u_0-u_2)},\\
\vphantom{\dac 1 2} X\equiv
\ddot{u}_1-\dot{u}_0\dot{u}_1+\dot{u}_1^2,& Y\equiv
\ddot{u}_2-\dot{u}_0\dot{u}_2+\dot{u}_2^2,\end{array}\end{equation}
\begin{equation}\label{signs2}\begin{array}{l}(p-m)_n\equiv(p-m)(p-m-1)(p-m-2)\ldots(p-n),\\ \vphantom{\dac 1 2}
(q-m)_n\equiv(q-m)(q-m-1)(q-m-2)\ldots(q-n),\end{array}\end{equation}
\begin{equation}\label{sigma}\sigma_p\equiv\dac{\widetilde{R}_p}{p(p-1)}\qquad
\sigma_q\equiv\dac{\widetilde{R}_q}{q(q-1)},\end{equation} where
$\widetilde{R}_p$ è $\widetilde{R}_q$ are internal curvatures of
$p$- and $q$-dimensional subspaces respectively,
$\widetilde{\Gamma}^i_{jk}$ and $\widetilde{\Gamma}^a_{bc}$ are
internal Christoffel symbols.

Now field equations are

\begin{equation}\label{eq0}\begin{array}{l}
\dac{1}{2}p_1A_p+\dac{1}{2}q_1A_q+pq\dot{u}_1\dot{u}_2+\dac1
2\dot{\varphi}^2-\dac{1}{2}e^{-2u_0}V(\varphi)+\vphantom{\dac{\dac{1}{2}}{2}}\dac1
2e^{-2u_0}\varepsilon(\varphi)\{p_3A_p^2+2p_1q_1A_pA_q+\\
\quad{}+q_3A_q^2+
4\dot{u}_1\dot{u}_2(p_2qA_p+pq_2A_q)+\vphantom{\dac{\dac{1}{2}}{2}}4p_1q_1\dot{u}_1^2\dot{u}_2^2\}+
2e^{-2u_0}\varepsilon'(\varphi)\dot{\varphi}\{A_p(p_2q\dot{u}_1+\\
\quad{}+qp_1\dot{u}_2)+
A_q(pq_1\dot{u}_1+\vphantom{\dac{\dac{1}{2}}{2}}
q_2\dot{u}_2)+2\dot{u}_1\dot{u}_2(p_1q\dot{u}_1+\vphantom{\dac{\dac{1}{2}}{2}}pq_1\dot{u}_2)\}=\dac{8\pi
G}{c^4}T_{00},\end{array}\end{equation}

\begin{equation}\label{eqij}\begin{array}{l}
e^{-2u_0}\{(1-p)X-qY-\dac{1}{2}(p-1)_2A_p-\dac 1 2 q_1
A_q-(p-1)q\dot{u}_1\dot{u}_2\}g_{ij}+\vphantom{\dac{\dac{1}{2}}{2}}\dac
1 2 e^{-2u_0}\dot{\varphi}^2g_{ij}+\\
\quad{}+\dac{1}{2}V(\varphi)g_{ij}-\dac{1}{2}\varepsilon(\varphi)
e^{-4u_0}g_{ij}\{(p-1)_4A_p^2+
\vphantom{\dac{\dac{1}{2}}{2}}4(p-1)_2q_1\dot{u}_1^2\dot{u}_2^2+4(p-1)_3A_pX+\\
\quad{}+\vphantom{\dac{\dac{1}{2}}{2}}
4(p-1)_3q\dot{u}_1\dot{u}_2A_p+8(p-1)q_1\dot{u}_1\dot{u}_2Y+
4(p-1)q_2\dot{u}_1\dot{u}_2A_q+\vphantom{\dac{\dac{1}{2}}{2}}
4(p-1)q_1A_qX+\\
\quad{}+8(p-1)_2q\dot{u}_1\dot{u}_2X+4(p-1)_2qA_pY+2(p-1)_2q_1A_pA_q+
\vphantom{\dac{\dac{1}{2}}{2}}q_3A_q^2+4q_2A_qY\}+\\ \quad{}+
2e^{-4u_0}g_{ij}\{-\vphantom{\dac{\dac{1}{2}}{2}}(\varepsilon''\dot{\varphi}^2+\varepsilon'\ddot{\varphi}-
\varepsilon'\dot{u}_0\dot{\varphi})[(p-1)_2A_p+q_1A_q+2(p-1)q\dot{u}_1\dot{u}_2]+\\
\quad{}+\vphantom{\dac{\dac{1}{2}}{2}}
\varepsilon'\dot{u}_1\dot{\varphi}[-2(p-1)_2X-2(p-1)qY-(p-1)_3A_p-(p-1)q_1A_q-\\
\quad{}- 2(p-1)_2q\dot{u}_1\dot{u}_2]+\vphantom{\dac{\dac{1}{2}}{2}}
\varepsilon'\dot{u}_2\dot{\varphi}[-2(p-1)qX-2q_1Y- (p-1)_2qA_p-\\
\quad{}-q_2A_q-\vphantom{\dac{\dac{1}{2}}{2}}
2(p-1)q_1\dot{u}_1\dot{u}_2]\}=\dac{8\pi G}{c^4}T_{ij},\end{array}
\end{equation}

\begin{equation}\label{eqab}\begin{array}{l}
e^{-2u_0}\{(1-q)Y-pX-\dac{1}{2}(q-1)_2A_q-\dac{1}{2}p_1A_p-(q-1)p\dot{u}_1\dot{u}_2\}g_{ab}+
\dac{1}{2}e^{-2u_0}\dot{\varphi}^2g_{ab}+\vphantom{\dac{\dac{1}{2}}{2}}\\
\quad{}+\dac{1}{2}V(\varphi)g_{ab}- \dac{1}{2}\varepsilon(\varphi)
e^{-4u_0}g_{ab}\{(q-1)_4A_q^2+4(q-1)_2p_1\dot{u}_1^2\dot{u}_2^2+
4(q-1)_3A_qY+\vphantom{\dac{\dac{1}{2}}{2}}\\ \quad{}+4(q-1)_3p
\dot{u}_1\dot{u}_2A_q+\vphantom{\dac{\dac{1}{2}}{2}}
8(q-1)p_1\dot{u}_1\dot{u}_2X+4(q-1)p_2\dot{u}_1\dot{u}_2A_p+
4(q-1)p_1A_pY+\\ \quad{}+
8(q-1)_2p\dot{u}_1\dot{u}_2Y+\vphantom{\dac{\dac{1}{2}}{2}}
4(q-1)_2pA_qX+2(q-1)_2p_1A_pA_q+p_3A_p^2+ 4p_2A_pX\}+\\
\quad{}+2e^{-4u_0}g_{ab}\{-(\varepsilon''\dot{\varphi}^2+\varepsilon'\ddot{\varphi}-
\varepsilon'\dot{u}_0\dot{\varphi})\vphantom{\dac{\dac{1}{2}}{2}}
[(q-1)_2A_q+p_1A_p+2(q-1)p\dot{u}_1\dot{u}_2]+\\ \quad{}+
\varepsilon'\dot{u}_2\dot{\varphi}[-2(q-1)_2Y-2(q-1)pX-\vphantom{\dac{\dac{1}{2}}{2}}
(q-1)_3A_q-(q-1)p_1A_p-\\ \quad{}-2(q-1)_2p\dot{u}_1\dot{u}_2]+
\varepsilon'\dot{u}_1\dot{\varphi}[-2(q-1)pY-2p_1X-
(q-1)_2pA_q-\vphantom{\dac{\dac{1}{2}}{2}}\\ \quad{}-p_2A_p-
2(q-1)p_1\dot{u}_1\dot{u}_2]\}=\vphantom{\dac{\dac{1}{2}}{2}}\dac{8\pi
G}{c^4}T_{ab},\end{array}\end{equation}

\begin{equation}\label{eqphi}\begin{array}{l}2[\ddot{\varphi}+(-\dot{u}_0+p\dot{u}_1+q\dot{u}_2)\dot{\varphi}]-
V'(\varphi)+\varepsilon'(\varphi)e^{-2u_0}
\left\{p_3A_p^2+2p_1q_1A_pA_q+q_3A_q^2+\vphantom{\dac 1 2} \right.\\
\quad{}+\left.\vphantom{\dac{\dac{1}{2}}{2}}
4\dot{u}_1\dot{u}_2(p_2qA_p+pq_2A_q)+4p_1q_1\dot{u}_1^2\dot{u}_2^2+
4pX[(p-1)_2A_p+q_1A_q+\right.\\
\quad{}+2(p-1)q\dot{u}_1\dot{u}_2]+\vphantom{\dac 1
2}\vphantom{\dac{\dac{1}{2}}{2}}\left.\vphantom{\dac 1 2}
4qY[p_1A_p+(q-1)_2A_q+2p(q-1)\dot{u}_1\dot{u}_2]\right\}=0.\end{array}\end{equation}

These equations are equivalent to those in \cite{bamba-guo-ohta} if
we substitute $g^{\mu\nu}\d_\mu\varphi\d_\nu\varphi$ by $-\frac 1 2
g^{\mu\nu}\d_\mu\varphi\d_\nu\varphi$ in Lagrangian and put
$\varepsilon(\varphi)=\alpha_2 e^{-\gamma\varphi}$, $V(\varphi)=0$,
$T_{\mu\nu}=0$.

Henceforth, we may put $u_0=0$ (for simplification), $p=3$ (to
identify $p$-subspace with visible space).

\subsection{Stationary solutions}

Let us now turn to find solutions of (\ref{eq0})--(\ref{eqphi})
under $p=3$, $u_0=0$. The simplest solutions are stationary ones.
Hence put
$$u_1=\const,\quad u_2=\const,\quad \varphi=\const.$$ Then
(\ref{signs1}) are $$A_p=\sigma_p e^{-2u_1},\quad A_q=\sigma_q
e^{-2u_2},\quad X=Y=0.$$

Consider space with homogeneous dust, i. e. $T_{00}\neq 0$ (other
$T_{\mu\nu}=0$).

After that, system (\ref{eq0})--(\ref{eqphi}) get the form of
algebraic equations
\begin{equation}\label{eq0-stat}3A_p+\dac{1}{2}q_1
A_q-\dac{1}{2}V(\varphi)+ \dac{1}{2}\varepsilon(\varphi)\{12q_1 A_p
A_q+q_3 A_q^2\}=\dac{8\pi
G}{c^4}T_{00};\vphantom{\dac{\dac{1}{2}}{2}}\end{equation}
\begin{equation}\label{eqij-stat}-A_p-\dac{1}{2}q_1
A_q+\dac{1}{2}V(\varphi)-\dac{1}{2}\varepsilon(\varphi)\{4q_1 A_q
A_p+q_3 A_q^2\}=0;\vphantom{\dac{\dac{1}{2}}{2}}\end{equation}
\begin{equation}\label{eqab-stat}-\dac{1}{2}(q-1)_2
A_q-3A_p+\dac{1}{2}V(\varphi)-\dac{1}{2}\varepsilon(\varphi)\{(q-1)_4
A_q^2+12(q-1)_2 A_p
A_q\}=0;\vphantom{\dac{\dac{1}{2}}{2}}\end{equation}
\begin{equation}\label{eqphi-stat}V'(\varphi)-\varepsilon'(\varphi)\{12q_1
A_p A_q+q_3 A_q^2\}=0.\vphantom{\dac{\dac{1}{2}}{2}}\end{equation}

From (\ref{eqphi-stat}) and (\ref{eq0-stat}) we have
\begin{equation}\label{eq0-phi}3A_p+\dac{1}{2}q_1
A_q=\dac{1}{2}V(\varphi)-\dac{1}{2}\varepsilon(\varphi)\dac{V'(\varphi)}{\varepsilon'(\varphi)}+
\dac{8\pi G}{c^4}T_{00}.\end{equation}

Try to find linear combination of
(\ref{eq0-stat})--(\ref{eqab-stat}) in order to cancel terms with
$\varepsilon(\varphi).$ Let $\alpha,\beta,\gamma$ are coefficients
for (\ref{eqij-stat}), (\ref{eqab-stat}) and (\ref{eq0-stat}) in
such a combination. Then we need $$\left\{\begin{array}{l}\alpha q_3+\beta (q-1)_4=\gamma q_3,\\
4\alpha q_1+12\beta (q-1)_2=12\gamma q_1.\end{array}\right.$$
Therefore
$$\gamma=\(1-\dac{1}{q}\)\beta,\quad\alpha=\dac{3}{q}\beta.$$
Now put $\beta=q.$ Then $\alpha=3,\quad \gamma=q-1.$

Now multiplying (\ref{eqij-stat}) by $3/(q+1)$, (\ref{eqab-stat}) by
$q/(q+1)$ and putting them together we have (taking
(\ref{eqphi-stat}) into account)
\begin{equation}\label{ij-ab-phi}3A_p+\dac{1}{2}q_1
A_q=\dac{q+3}{2(q+1)}V(\varphi)-\dac{q-1}{2(q+1)}\varepsilon(\varphi)
\dac{V'(\varphi)}{\varepsilon'(\varphi)}.\end{equation} From this
and (\ref{eq0-phi}) one can get
\begin{equation}\label{0-ij-ab-phi}\dac{1}{q+1}V(\varphi)+
\dac{1}{q+1}\varepsilon(\varphi)\dac{V'(\varphi)}{\varepsilon'(\varphi)}-\dac{8\pi
G}{c^4}T_{00}=0.\end{equation}

Put now $$V(\varphi)=ae^{-\alpha\varphi},\quad
\varepsilon(\varphi)=be^{-\beta\varphi}.$$ Then (\ref{0-ij-ab-phi})
get the form
$$\dac{1}{q+1}\(1+\dac{\alpha}{\beta}\)ae^{-\alpha\varphi}-
\dac{8\pi G}{c^4}T_{00}=0.$$

It is easy to see:
\begin{equation}\label{phi-state}\varphi=-\dac{1}{\alpha}\ln\left\{\dac{\beta(q+1)}{a(\alpha+\beta)}\cdot
\dac{8\pi G}{c^4}T_{00}\right\}.\end{equation}

Plugging this $\varphi$ into (\ref{eq0-phi}) we have:
\begin{equation}\label{eq0-varphi}3A_p+\dac{1}{2}q_1 A_q=\dac{4\pi
G}{c^4}T_{00}\dac{(1-q)\alpha+(q+3)\beta}{\alpha+\beta}.\end{equation}
Plugging $A_p$ derived from that into (\ref{eqphi-stat}) we can get
$$A_q^2
q_1\{(q-2)(q-3)-2q(q-1)\}+2q_1\varkappa\dac{(1-q)\alpha+(q+3)\beta}{\alpha+\beta}A_q=
\dac{\alpha a}{\beta
b}\left\{\dac{\beta(q+1)}{a(\alpha+\beta)}\varkappa\right\}^{\frac{\alpha-\beta}{\alpha}},$$
where $$\varkappa\equiv \dac{8\pi G}{c^4}T_{00}.$$

That is quadratic equation on $A_q,$ which solutions are
\begin{equation}\label{Aq}A_q=\dac{-2q_1\varkappa
\dac{(1-q)\alpha+(q+3)\beta}{\alpha+\beta}\pm\sqrt{D}}{2(q_3-2q_1^2)},\end{equation}
where
\begin{equation}\label{D}D=\(2q_1\varkappa\dac{(1-q)\alpha+(q+3)\beta}{\alpha+\beta}\)^2+
4q_1\{(q-2)(q-3)-2q(q-1)\}\dac{\alpha a}{\beta
b}\left\{\dac{\beta(q+1)}{a(\alpha+\beta)}\varkappa\right\}^{\frac{\alpha-\beta}{\alpha}}.\end{equation}

Then, taking (\ref{eq0-varphi}) into account, we obtain
\begin{equation}\label{Ap}A_p=\dac{1}{6}\varkappa\dac{(1-q)\alpha+(q+3)\beta}{\alpha+\beta}-
\dac{q_1}{6}A_q.\end{equation}

Finally, \begin{equation}\label{u1u2-state}u_1=-\dac{1}{2}\ln
\dac{A_p}{\sigma_p},\quad u_2=-\dac{1}{2}\ln
\dac{A_q}{\sigma_q}\end{equation} and $\varphi$ is
(\ref{phi-state}).

It should be noted that the following constraints was applied:
$$T_{00}\neq 0,\quad a\neq 0,\quad b\neq 0,\quad \alpha\neq 0,\quad
\beta\neq 0,$$
$$\alpha\neq-\beta, \quad q\neq 0, \quad q\neq 2,\quad q\neq 4,\quad\sigma_p\neq
0,\quad \sigma_q\neq 0.$$

It is easy to derive solution for (3+1)-dimensional space with
perfect fluid of arbitrary equation of state parameter. At that we
should not specify $V(\varphi)$ and $\varepsilon(\varphi)$ because
of $\varepsilon(\varphi)$ do not participate in equations and
$V(\varphi)$ is specified from those. Dilaton also do not contribute
in equations, therefore we should solve just Einstein equations with
cosmological constant. Solution is $$V=(1+3w)\dac{8\pi
G}{c^4}T_{00},\qquad u_1=-\dac{1}{2}\ln\(\dac{1+w}{2\sigma_p}
\dac{8\pi G}{c^4}T_{00}\).$$ Here $\omega$ is equation of state
parameter ($p=w\epsilon$, $p$ is pressure, $\epsilon\equiv T_{00}$
is energy density). A particular case  (when $w=0$) was derived
by A.  Einstein in 1917 \cite{einstein1917}.

\subsection{Exponential solutions}\label{exp}

For the dynamical solutions we need to do further simplification of
(\ref{eq0})--(\ref{eqphi}). Therefore, in addition to $p=3$ and
$u_0=0$, put $\sigma_p=\sigma_q=0$ i. e. subspaces are flat. In
addition to simplicity, such a condition is caused by Cosmic
Microwave Background observations \cite{jaffe,spergel} indicate the
flatness of visible subspace. For extra subspace $\sigma_q=0$ is
only a simplification.

After that, equations (\ref{eq0})--(\ref{eqphi}) are
\begin{equation}\label{eq0-plane}\begin{array}{l}3\dot{u}_1^2+\dac{1}{2}q_1\dot{u}_2^2+
3q\dot{u}_1\dot{u}_2+\dac{1}{2}\dot{\varphi}^2-\dac{1}{2}V(\varphi)+
\dac{1}{2}\varepsilon(\varphi)\{36q_1\dot{u}_1^2\dot{u}_2^2+q_3\dot{u}_2^4
+12\dot{u}_1\dot{u}_2(2q\dot{u}_1^2 +q_2\dot{u}_2^2)
\}+\vphantom{\dac{\dac{1}{2}}{2}}\\
\quad{}+2\varepsilon'\dot{\varphi}\{6q\dot{u}_1^3+18q\dot{u}_1^2\dot{u}_2
+9q_1\dot{u}_1\dot{u}_2^2
+q_2\dot{u}_2^3\vphantom{\dac{\dac{1}{2}}{2}}\}=\dac{8\pi
G}{c^4}T_{00}\vphantom{\dac{\dac{1}{2}}{2}}.\end{array}\end{equation}

\begin{equation}\label{eqij-plane}\begin{array}{l}\{-2\ddot{u}_1-q\ddot{u}_2-3\dot{u}_1^2
-\dac{1}{2}q(q+1)\dot{u}_2^2
-2q\dot{u}_1\dot{u}_2\}g_{ij}+\vphantom{\dac{\dac{1}{2}}{2}}
\dac{1}{2}\dot{\varphi}^2g_{ij}+\dac{1}{2}V(\varphi)g_{ij}-\\
\quad{}-
\dac{1}{2}\varepsilon(\varphi)g_{ij}\vphantom{\dac{\dac{1}{2}}{2}}\{8\ddot{u}_1\dot{u}_2
(q_1\dot{u}_2+2q\dot{u}_1)+
4\ddot{u}_2(4q_1\dot{u}_1\dot{u}_2+2q\dot{u}_1^2+q_2\dot{u}_2^2)+
(q+1)q_2\dot{u}_2^4+\\
\quad{}+4q(5q-3)\dot{u}_1^2\dot{u}_2^2+8qq_1\dot{u}_1\dot{u}_2^3+
16q\dot{u}_1^3\dot{u}_2\}+2g_{ij}\{-(\varepsilon''\dot{\varphi}^2+
\varepsilon'\ddot{\varphi})[2\dot{u}_1^2 +q_1\dot{u}_2^2+\vphantom{\dac{\dac{1}{2}}{2}}\\
\quad{}+ 4q\dot{u}_1\dot{u}_2]+
\varepsilon'\dot{u}_1\dot{\varphi}[-4\ddot{u}_1-4q\ddot{u}_2-4\dot{u}_1^2-
2(q+1)q\dot{u}_2^2-\vphantom{\dac{\dac{1}{2}}{2}}
4q\dot{u}_1\dot{u}_2]+\\ \quad{}+
\varepsilon'\dot{u}_2\dot{\varphi}[-4q\ddot{u}_1-2q_1\ddot{u}_2-6q\dot{u}_1^2
-qq_1\dot{u}_2^2-4q_1\dot{u}_1\dot{u}_2]\}=
\vphantom{\dac{\dac{1}{2}}{2}}\dac{8\pi
G}{c^4}T_{ij}.\end{array}\end{equation}

\begin{equation}\label{eqab-plane}\begin{array}{l}\{-3\ddot{u}_1-
(q-1)\ddot{u}_2-6\dot{u}_1^2-\dac{1}{2}q_1\dot{u}_2^2
-3(q-1)\dot{u}_1\dot{u}_2\}g_{ab}+
\dac{1}{2}\dot{\varphi}^2g_{ab}+\dac{1}{2}V(\varphi)g_{ab}-\vphantom{\dac{\dac{1}{2}}{2}}\\
\quad{}-\vphantom{\dac{\dac{1}{2}}{2}}
\dac{1}{2}\varepsilon(\varphi)g_{ab}\{12\ddot{u}_1(2\dot{u}_1^2+(q-1)_2\dot{u}_2^2
+4(q-1)\dot{u}_1\dot{u}_2)+4\ddot{u}_2(6(q-1)_2\dot{u}_1\dot{u}_2+\\
\quad{}+(q-1)_3\dot{u}_2^2+
6(q-1)\dot{u}_1^2)+24\dot{u}_1^4+\vphantom{\dac{\dac{1}{2}}{2}}
q_3\dot{u}_2^4+\vphantom{\dac{\dac{1}{2}}{2}}
24(q-1)(2q-3)\dot{u}_1^2\dot{u}_2^2+\\ \quad{}+
72(q-1)\dot{u}_1^3\dot{u}_2+12(q-1)(q-1)_2\dot{u}_1\dot{u}_2^3\}+
2g_{ab}\{-(\varepsilon''\dot{\varphi}^2+\vphantom{\dac{\dac{1}{2}}{2}}
\varepsilon'\ddot{\varphi})[6\dot{u}_1^2 +\\ \quad{}+
\vphantom{\dac{\dac{1}{2}}{2}}
(q-1)_2\dot{u}_2^2+6(q-1)\dot{u}_1\dot{u}_2]\vphantom{\dac{\dac{1}{2}}{2}}
+\varepsilon'\dot{u}_2\dot{\varphi}[-6(q-1)
\ddot{u}_1-2(q-1)_2\ddot{u}_2-\\
\quad{}-12(q-1)\dot{u}_1^2-(q-1)(q-1)_2\dot{u}_2^2-
6(q-1)_2\dot{u}_1\dot{u}_2] \vphantom{\dac{\dac{1}{2}}{2}}+
\varepsilon'\dot{u}_1\dot{\varphi}[-12\ddot{u}_1-\\
\quad{}-6(q-1)\ddot{u}_2-18\dot{u}_1^2
-3q_1\dot{u}_2^2-12(q-1)\dot{u}_1\dot{u}_2]\}=\dac{8\pi
G}{c^4}T_{ab}.\vphantom{\dac{\dac{1}{2}}{2}}\end{array}\end{equation}

\begin{equation}\label{eqphi-plane}\begin{array}{l}-2\ddot{\varphi}-2(3\dot{u}_1+q\dot{u}_2)\dot{\varphi}+
V'(\varphi)-\varepsilon'(\varphi) \{12\ddot{u}_1[2\dot{u}_1^2+
q_1\dot{u}_2^2+4q\dot{u}_1\dot{u}_2]+4q\ddot{u}_2[6\dot{u}_1^2+(q-1)_2\dot{u}_2^2+\\
\quad{}+
6(q-1)\dot{u}_1\dot{u}_2]+24\dot{u}_1^4+\vphantom{\dac{\dac{1}{2}}{2}}
(q+1)q_2\dot{u}_2^4+
24q(2q-1)\dot{u}_1^2\dot{u}_2^2+72q\dot{u}_1^3\dot{u}_2+12qq_1\dot{u}_1\dot{u}_2^3
\}=0.\vphantom{\dac{\dac{1}{2}}{2}}\end{array}\end{equation}

Find at first solutions without dilaton, without matter and with
constant Hubble parameters:
\begin{equation}\label{hubble=const}\varphi=0,\qquad V(\varphi)=0,\qquad
T_{\mu\nu}=0,\qquad\dot{u}_1=\const,\qquad
\dot{u}_2=\const.\end{equation} Then system
(\ref{eq0-plane})--(\ref{eqab-plane}) will be a system of algebraic
equations for which we have found two analytical solutions for
arbitrary $q$ and negative $\varepsilon$:
\begin{equation}\label{exponent-general}\dot{u}_1=\dot{u}_2=\pm\dac{1}{\sqrt{-q(q+1)\varepsilon}}.\end{equation}

Also  particular cases from $q=1$ to $q=22$
(22+4=26 is required for the bosonic strings, and in the case of
$q=0$ Lovelock gravity is just Einstein gravity) have been studied, but solutions
different from (\ref{exponent-general}) have been obtained only for
$q=3$ (i. e. for 7-dimensional space just as in section 1):
\begin{equation}\label{exponent3-1}\dot{u}_1=-\zeta_1\sqrt{\dac{3+\zeta_2\sqrt{5}}{2\varepsilon}}\cdot
\dac{575+257\zeta_2\sqrt{5}}{3010+1346\zeta_2\sqrt{5}},\qquad
\dot{u}_2=\zeta_1\sqrt{\dac{3+\zeta_2\sqrt{5}}{8\varepsilon}},\end{equation}
and
\begin{equation}\label{exponent3-2}\dot{u}_1=\zeta_1\sqrt{\dac{3+\zeta_2\sqrt{5}}{8\varepsilon}},\qquad
\dot{u}_2=-\zeta_1\sqrt{\dac{3+\zeta_2\sqrt{5}}{2\varepsilon}}\cdot
\dac{575+257\zeta_2\sqrt{5}}{3010+1346\zeta_2\sqrt{5}},\end{equation}
where constants $\zeta_1$ and $\zeta_2$ take values of $+1$ and $-1$
independently from each other, and $\varepsilon>0$.

Therefore scale factors are $$a(t)\equiv e^{u_1}=a_0 e^{\dot{u}_1
t},\qquad b(t)\equiv e^{u_2}=b_0 e^{\dot{u}_2 t}.$$

It is clear that solutions (\ref{exponent-general}) are not useful
for us by the following cause: when visible subspace expands, extra
subspace expands too, then extra subspace must be visible in this
case. But solutions (\ref{exponent3-1}) for $\zeta_1=-1$ and
(\ref{exponent3-2}) for $\zeta_1=+1$ satisfy our purpose.

Now let us try to obtain exponential solutions in the presence of
perfect fluid. For that substitute conditions (\ref{hubble=const})
by
\begin{equation}\label{h=c-matter}\varphi=0,\quad V(\varphi)=0,\quad
T_{00}=\epsilon,\quad T_{ij}=w\epsilon g_{ij},\quad T_{ab}=w\epsilon
g_{ab},\quad\dot{u}_1=\const,\quad \dot{u}_2=\const.\end{equation}
After plugging those into (\ref{eq0-plane})--(\ref{eqab-plane}) and
subtracting factors $g_{ij}$ and $g_{ab}$ we see that left-hand
sides of equations are independent of time. Hence the right-hand
sides also must be constant.

From 00-component of local conservation law for energy-momentum
tensor ($\nabla^\mu T_{\mu 0}=0$) one can obtain (taking
(\ref{h=c-matter}) into account)  $$\epsilon=\epsilon_0
\exp[-(1+w)(3\dot{u}_1+q\dot{u}_2)t].$$ Therefore $\epsilon=\const$
under at least one of a two conditions (here $H\equiv\dot{u}_1$,
$h\equiv\dot{u}_2$):
\begin{enumerate}
\item $w=-1$;
\item $h=-\dac{3}{q}H.$
\end{enumerate}

In the first case matter can be described by cosmological constant,
in the second one comoving bulk is constant. In the latter case
equations (\ref{eq0-plane})--(\ref{eqab-plane}), as equations on
$H,\epsilon,w$, have two solutions:

\begin{enumerate}
\item $$H=0,\qquad \epsilon=0,\qquad w \mbox{ is arbitrary},$$ i. e. flat
space with Lorenz metric.
\item \begin{equation}\label{exp-any-matter-2}\begin{array}{ll}H \mbox{ is arbitrary},\qquad &\epsilon=\dac{3c^4
H^2(-q^3-3q^2+3\varepsilon H^2 q^3+54\varepsilon H^2
q^2+81\varepsilon H^2 q -162\varepsilon H^2)}{16\pi G q^3},\\
&w=\dac{\varepsilon H^2 q^2 -q^2+15\varepsilon H^2 q-18\varepsilon
H^2}{-q^2+3\varepsilon H^2 q^2+45\varepsilon H^2 q-54\varepsilon
H^2},\vphantom{\dac{\dac 1 2}{1}}\end{array}\end{equation} i. e. one
can obtain any value for $H$ by matching energy density $\epsilon$
and EoS parameter $w$. It is clear that $h<0$ if $H>0$, that's why
this solution satisfies all requirements. Finally, such a solution
describes anisotropic expansion of the Universe with matter which
not tends to isotropization. In Einstein gravity it is possible only
for maximally stiff fluid: $w=1$.
\end{enumerate}

Now turn to the cosmological constant case: $w=-1$. Then it is
possible to consider (\ref{eq0-plane})--(\ref{eqab-plane}) as
equations on $H$, $h$ and $\epsilon$. These have the following
solutions:
\begin{enumerate}
\item \begin{equation}\label{exp-cosm-const-1}\begin{array}{l}H=h \mbox{ and is arbitrary},\\
\epsilon=\dac{c^4 h^2(6+q^2+5q+\varepsilon q^4 h^2+6\varepsilon q^3
h^2+11h^2\varepsilon q^2+6h^2 \varepsilon q)}{16\pi
G};\vphantom{\dac{\dac 1 2}{1}}\end{array}\end{equation}
\item \begin{equation}\label{exp-cosm-const-2}\begin{array}{l}h \mbox{ is arbitrary},\quad H=-(q-1)h\pm\dac{\sqrt{2\varepsilon^2 h^2
q(q-1)-\varepsilon}}{2\varepsilon}\\ \quad{}
\epsilon=\dac{c^4}{32\varepsilon\pi G}\(-96H\varepsilon^2 q^2
h^3+48H\varepsilon^2 q h^3-24H\varepsilon q h+48H\varepsilon^2
q^3 h^3+24Hh\varepsilon+2h^2\varepsilon q^2+\right.\\
\left.\quad\qquad{}+10h^2 \varepsilon q+14\varepsilon^2 q^4
h^4-60\varepsilon^2 q^3 h^4+82\varepsilon^2 q^2 h^4-36\varepsilon^2
q h^4-3-12h^2\varepsilon\)\vphantom{\dac 1
2}.\end{array}\end{equation}
\end{enumerate}

It is evident that the first solution is unsatisfactory. The second
one is adequate under $h<0$, $H>0$. Such conditions are fulfilled in
three cases:
\begin{enumerate}
\item $\varepsilon>0,\quad h<-\dac{1}{\sqrt{2\varepsilon q(q-1)}},$ sign in expression for $H$ is arbitrary.
\item $\varepsilon<0$, sign is "$-$".
\item $\varepsilon<0,\quad h<-\dac{1}{\sqrt{-2\varepsilon(q-2)(q-1)}}$, sign is "$+$".
\end{enumerate}

Here we also should emphasize an existence of solutions with
matter which do not tend to isotropization.

\subsection{Exponent-of-exponent form solutions}

Try to obtain solutions with a dynamical dilaton. For that purpose
consider equations (\ref{eq0-plane})--(\ref{eqphi-plane}) and notice
that functions $u_1(t)$ and $u_2(t)$ make contribution only through
derivatives $\dot{u}_1$, $\dot{u}_2$, $\ddot{u}_1$, $\ddot{u}_2$,
but $\varphi(t)$ participate explicitly. To find solutions with
constant derivatives let's eliminate $\varphi(t)$ by introducing new
time variable. At first we put
\begin{equation}\label{epsilonV}\varepsilon(\varphi)=\beta
e^{-\gamma\varphi},\quad V(\varphi)=\alpha e^{\gamma\varphi},\quad
T_{\mu\nu}=0.\end{equation}

Now turn from time $t$ to new variable $\tau$:
$$\d_\tau=e^{-\gamma\varphi/2}\d_t.$$ Derivatives with respect to
$\tau$ will be denoted by the prime $'$. After such a substitution
and putting $u_1''=u_2''=\varphi''=0$ one can obtain
\begin{equation}\label{eq0-exp-exp}\begin{array}{l} 3{u_1'}^2+\dac{1}{2}q_1 {u_2'}^2+3q u_1'u_2'+
\dac{1}{2}{\varphi'}^2-\dac{1}{2}\alpha+\dac{1}{2}\beta\{36q_1{u_1'}^2{u_2'}^2+
q_3{u_2'}^4+\\
\quad{}+12u_1'u_2'(2q{u_1'}^2+q_2{u_2'}^2)\}-\vphantom{\dac{\dac{1}{2}}{2}}
2\beta\gamma\varphi'\{18q{u_1'}^2 u_2'+9q_1 u_1'{u_2'}^2+
6q{u_1'}^3+q_2{u_2'}^3\}=0;\vphantom{\dac{\dac{1}{2}}{2}}\end{array}
\end{equation}

\begin{equation}\label{eqij-exp-exp}\begin{array}{l}-3{u_1'}^2-\dac{1}{2}(q+1)q{u_2'}^2-
\gamma\varphi'\(u_1'+\dac{q}{2}u_2'\)-2qu_1'u_2'+\dac{1}{2}{\varphi'}^2+
\dac{\alpha}{2}-\dac{1}{2}\beta\{-4q(3q+1)\gamma\varphi'u_1'{u_2'}^2-\\
\quad{}-
28q\gamma\varphi'{u_1'}^2u_2'-\vphantom{\dac{\dac{1}{2}}{2}}
2(q+2)q_1\gamma\varphi'{u_2'}^3-4\gamma^2{\varphi'}^2{u_1'}^2-2q_1\gamma^2{\varphi'}^2{u_2'}^2+
4q(5q-3){u_1'}^2{u_2'}^2+\vphantom{\dac{\dac{1}{2}}{2}}\\
\quad{}+
8qq_1u_1'{u_2'}^3+16q{u_1'}^3u_2'+(q+1)q_2{u_2'}^4-16\gamma\varphi'{u_1'}^3-
8q\gamma^2{\varphi'}^2u_1'u_2'\}=0;\vphantom{\dac{\dac{1}{2}}{2}}\end{array}\end{equation}

\begin{equation}\label{eqab-exp-exp}\begin{array}{l}-\dac{1}{2}q_1{u_2'}^2-6{u_1'}^2-\dac{q-1}{2}\gamma\varphi'u_2'-
\dac{3}{2}\gamma\varphi'u_1'-3(q-1)u_1'u_2'+\dac{1}{2}{\varphi'}^2+\dac{\alpha}{2}-{}\\
\quad{}-\vphantom{\dac{\dac{1}{2}}{2}}\dac{1}{2}\beta
\{-2(q+1)(q-1)_2\gamma\varphi'{u_2'}^3-
6(q-1)(3q-2)\gamma\varphi'u_1'{u_2'}^2-
60(q-1)\gamma\varphi'{u_1'}^2u_2'-\\
\quad{}-\vphantom{\dac{\dac{1}{2}}{2}}60\gamma\varphi'{u_1'}^3-2(q-1)_2\gamma^2{\varphi'}^2{u_2'}^2-12\gamma^2{\varphi'}^2{u_1'}^2+
q_3{u_2'}^4+\vphantom{\dac{\dac{1}{2}}{2}}24(q-1)(2q-3){u_1'}^2{u_2'}^2+\\
\quad{}+12(q-1)(q-1)_2u_1'{u_2'}^3+72(q-1){u_1'}^3u_2'+
\vphantom{\dac{\dac{1}{2}}{2}}24{u_1'}^4-12(q-1)\gamma^2{\varphi'}^2u_1'u_2'\}=0;\end{array}\end{equation}

\begin{equation}\label{eqphi-exp-exp}\begin{array}{l}-\gamma {\varphi'}^2-2(3u_1'+q u_2')\varphi'+
\beta\gamma
\{12q(4q-2){u_1'}^2{u_2'}^2+(q+1)q_2{u_2'}^4+72q{u_1'}^3u_2'+
12qq_1u_1'{u_2'}^3+\\
\quad{}+\vphantom{\dac{\dac{1}{2}}{2}}18q_1\gamma\varphi'u_1'{u_2'}^2+36q\gamma\varphi'{u_1'}^2u_2'+
12\gamma\varphi'{u_1'}^3+24{u_1'}^4+2q_2\gamma\varphi'{u_2'}^3
\}+\alpha\gamma=0.\vphantom{\dac{\dac{1}{2}}{2}}\end{array}\end{equation}

Now assume that we have obtained some quantities $u_1', u_2',
\varphi'$ which satisfy these equations. What should they be to
describe accelerating expansion of visible subspace and contraction
of extra one? It is easy to see that the scale factor of visible
subspace would be
\begin{equation}\label{scale-1}a(t)\equiv e^{u_1}=a_0
\exp\left\{\dac{2u_1'c_0}{\varphi'}e^{\varphi't/2}\right\},\end{equation}
where $a_0, c_0$ are arbitrary positive constants. Then its first
and second derivatives with respect to time $t$ would be
$$\begin{array}{l}\dot{a}(t)=a_0 u_1' c_0
e^{\varphi't/2}\exp\left\{\dac{2u_1'c_0}{\varphi'}e^{\varphi't/2}\right\},\\
\vphantom{\dac{\dac{1}{2}}{2}} \ddot{a}(t)=\dac{1}{2}a_0 u_1' c_0
\varphi'e^{\varphi't/2}\exp\left\{\dac{2u_1'c_0}{\varphi'}e^{\varphi't/2}\right\}+
a_0 {u_1'}^2 c_0^2
e^{\varphi't}\exp\left\{\dac{2u_1'c_0}{\varphi'}e^{\varphi't/2}\right\}.\end{array}$$
It is clear that both derivatives would be positive (as need for
accelerating expansion) if $u_1'>0, \varphi'>0$ (the latter is not
necessary but is sufficient).

By the same manner we obtain the scale factor for extra dimensions:
\begin{equation}\label{scale-2}b(t)\equiv e^{u_2}=b_0
\exp\left\{\frac{2u_2'c_0}{\varphi'}e^{\varphi't/2}\right\}\end{equation}
($b_0$ is positive constant) and its first derivative:
$$\dot{b}(t)=b_0 u_2' c_0
e^{\varphi't/2}\exp\left\{\frac{2u_2'c_0}{\varphi'}e^{\varphi't/2}\right\},$$
which would be negative (as need for contraction) under $u_2'<0$.

Therefore it is necessary to find solutions of
(\ref{eq0-exp-exp})--(\ref{eqphi-exp-exp}) satisfied conditions
$$u_1'>0,\quad u_2'<0,\quad \varphi'>0.$$ Numerical calculations
give us solutions for different dimensions from $q=1$ to $q=20$. For
example,
$$\begin{array}{lllll}q=1,\quad& \alpha=1,\beta=1,\gamma=1,\quad& \varphi'=0.383,\quad&
u_1'=0.378,\quad& u_2'=-1.32.\vphantom{dac{1}{2}}\\
q=2,\quad& \alpha=0.01,\beta=100,\gamma=0.01,\quad&
\varphi'=5.09\cdot10^{-5},\quad& u_1'=0.0175,\quad& u_2'=-0.0803.\vphantom{dac{1}{2}}\\
q=3,\quad& \alpha=0.001,\beta=100,\gamma=0.001,\quad&
\varphi'=2.33\cdot10^{-4},\quad& u_1'=0.0306,\quad&
u_2'=-0.0788.\vphantom{dac{1}{2}}\end{array}$$

Finally, field equations in the case of flat subspaces without
matter and with $V(\varphi)$ and $\varepsilon(\varphi)$ in the form
of (\ref{epsilonV}) have exponent-of-exponent form solutions
(\ref{scale-1}), (\ref{scale-2}) with abovementioned parameters.

\subsection{Power-law solutions}

Now consider space with dust-like matter: $$T_{00}\neq
0,\qquad\mbox{other } T_{\mu\nu}=0.$$ And try to obtain solutions of
system (\ref{eq0-plane})--(\ref{eqphi-plane}) with scale factors of
power-law form:
\begin{equation}\label{ab-power}a(t)\equiv
e^{u_1}=\(\dac{t}{t_1}\)^n,\qquad b(t)\equiv
e^{u_2}=\(\dac{t}{t_2}\)^m.\end{equation}

Then all Einstein terms in left-hand sides of equations
(\ref{eq0-plane})--(\ref{eqab-plane}) are proportional to $1/t^2$
and Gauss-Bonnet terms are proportional to $1/t^4$. Hence solutions
with scale factors (\ref{ab-power}) are possible if
\begin{equation}\label{T00-prop}T_{00}\propto
\dac{1}{t^2},\end{equation}
\begin{equation}\label{kinetic-prop}\dot{\varphi}^2\propto
\dac{1}{t^2},\end{equation}
\begin{equation}\label{epsilon-prop}\varepsilon(\varphi)\propto
t^2,\end{equation}
\begin{equation}\label{V-prop}V(\varphi)\propto
\dac{1}{t^2},\end{equation} Consider these conditions one by one.

It is possible to derive time dependence of energy density from
conservation law of energy-momentum tensor:
$$T^{00}=\const\cdot t^{-3n-qm}.$$ From comparison this expression
with (\ref{T00-prop}) we have $$m=\dac{2-3n}{q}.$$ Note that under
this condition extra subspace contracts ($m<0$) if visible subspace
expanses accelerative ($n>1$) (but we haven't obtain such a
solution, see below). For (3+1)-dimensional space condition
(\ref{T00-prop}) leads to $n=2/3$ i. e. to Friedmann solution.

From condition (\ref{kinetic-prop}) it is easy to obtain
$\varphi(t)=\psi \ln(t/t_3),$ where $\psi$, $t_3$ are arbitrary
constants ($t_3>0$). In order to avoid unnecessary complication put
$\psi=1$. Therefore $$\varphi(t)=\ln\(\dac{t}{t_3}\).$$

From comparison that with (\ref{epsilon-prop}), (\ref{V-prop}) we
see:
$$V(\varphi)=\tilde{\alpha}e^{-2\varphi},\qquad
\varepsilon(\varphi)=\tilde{\beta}e^{2\varphi},$$ where
$\tilde{\alpha}$, $\tilde{\beta}$ are constants.

Plugging all those into equations
(\ref{eq0-plane})--(\ref{eqphi-plane}) and putting
$$\tilde{\alpha}\rightarrow\alpha=\tilde{\alpha}t_3^2, \qquad
\tilde{\beta}\rightarrow\beta=\tilde{\beta}/t_3^2,$$ one can obtain
a system of algebraic equations on $n$ with parameters $\alpha$,
$\beta$, $q$ and $\varkappa\equiv\dac{8\pi G}{c^4}T_{00}(t_0)t_0^2$,
where $T_{00}(t_0)$ is energy density at some time moment $t_0$.
Such a system of equations has solutions not at all values of
parameters. Considering dimensions from $q=1$ to $q=22$ we have
obtained a set of solutions with arbitrary $\varkappa$. Here
$\alpha$ and $\beta$ are functions of $\varkappa$ and $0<n=m<1$ i.
e. visible and extra subspaces expands with deceleration (and with
the same velocity). In another set of solutions $\varkappa$
possesses fixed values. Here in two cases $n=m$ and in another
cases $n\neq m$. These are solutions for $q=1,2,6,9,12$:
$$q=1,\qquad \varkappa\mbox{ is arbitrary},\qquad
\alpha=\dac{3}{7}\varkappa-\dac{29}{14},\qquad
\beta=\dac{2}{21}\varkappa-\dac{5}{21},\qquad n=m=\dac{1}{2}.$$

$$q=2,\qquad \varkappa=\dac{53}{40},\qquad
\alpha=-\dac{973}{640},\qquad \beta=-\dac{25}{272},\qquad
n=\dac{3}{10},\qquad m=\dac{11}{20}.$$

$$q=6,\qquad \varkappa\mbox{ is arbitrary},\qquad
\alpha=\dac{112}{253}\varkappa-\dac{5077}{2277},\qquad
\beta=\dac{729}{16192}\varkappa-\dac{2025}{16192},\qquad
n=m=\dac{2}{9}.$$

$$q=6,\qquad \varkappa=\dac{51397}{39528},\qquad
\alpha=-\dac{21035}{13176},\qquad \beta=-\dac{243}{3904},\qquad
n=\dac{8}{27},\qquad m=\dac{5}{27}.$$

$$q=9,\qquad \varkappa=\dac{17}{6},\qquad \alpha=-1,\qquad
\beta=0,\qquad n=m=\dac{1}{6}.$$

$$q=12,\qquad \varkappa=\dac{43}{15},\qquad \alpha=-1,\qquad
\beta=0,\qquad n=m=\dac{2}{15}.$$

Note that in the last two cases $\beta=0$, therefore these solutions
are solutions in Brans-Dicke theory (i. e. theory with Lagrangian
(\ref{EGBd-lagr}) without Gauss-Bonnet term).

In all obtained power-law solutions $0<n<1$, $0<m<1$, that's why
such solutions don't describe accelerated expansion of visible space
or contraction of extra dimensions. However solutions with $n\neq m$
are interesting for another cause. In Einstein theory there is no
anisotropic power-law solution in the presence of dust. However, in
Einstein-Gauss-Bonnet theory with dilaton that is possible.

\section*{Conclusion}

Different variants of Lovelock gravity with dilaton were considered
in $D$-dimensional space with two maximally symmetric subspaces:
$3$-dimensional and $(D-4)$-dimensional. Absence of matter and
existence of perfect fluid were investigated. We have several types of obtained solutions:

\begin{enumerate}
\item Stationary.
\item Power-law.
\item Exponential.
\item Exponent-of-exponent form solutions.
\end{enumerate}
Among the last two forms solutions which describe accelerating
expansion of 3-dimensional subspace and contraction of
$(D-4)$-dimensional one were elected. Unobservability of the latter
subspace was justified on the basis of Kaluza-Klein approach. Also a
set of anisotropic solutions which do not tend to isotropization in
the presence of matter, in contrast to Einstein gravity, have been
obtained. Such a possibility is of importance because it allows us
to assume that extra dimensions become small during the Universe
evolution. This issue we are going to investigate in more detail in
another work. Moreover, it would be interesting to extend the
results of this work for account of third-order Lovelock terms. This
will be done elsewhere.

Studying of future singularities in
such models would also be important. For 4-dimensional modified gravities this problem was
considered in \cite{singularities1, singularities2}.

Unfortunately most of solutions describe only flat maximally
symmetric subspaces. For curved subspaces there are only stationary
solutions. Those are of interest as exact solutions of very
complicated equations and as possible basis for numerical dynamical
solutions in the case of curved subspaces.

\section*{Acknowledgments}

This work is partially supported by RFBR grant 06-01-00609 and by RF
Presidential grant for LSS 2553.2008.2. I.V.K. is supported by
grants of Tomsk State University academic council and of "Dynasty"
foundation in the frameworks of International Center for Fundamental
Physics in Moscow. The authors are grateful to S. D. Odintsov, K. E.
Osetrin and A. V. Toporensky for useful discussions.

\end{document}